\documentclass{elsart}
\usepackage{graphicx}
\usepackage{dcolumn}
\usepackage{bm}
\usepackage{amsmath}
\usepackage{amsfonts}
\usepackage{amssymb}

\newcommand{\ket}[1]{|#1\rangle}

\begin{document}
\begin{frontmatter}
\title{Generalized Quantum Walk \\in Momentum Space}
\author{A. Romanelli, }
\author{A. Auyuanet\thanksref{PEPE}},
\thanks[PEPE]{Corresponding author. \textit
{E-mail address: auyuanet@fing.edu.uy}}
\author{R. Siri, }
\author{G. Abal, }
\author{and R. Donangelo\thanksref{UFRJ}}
\address{Instituto de F\'{\i}sica, Facultad de Ingenier\'{\i}a\\
Universidad de la Rep\'ublica\\ C.C. 30, C.P. 11000, Montevideo, Uruguay}
\thanks[UFRJ]{Permanent address: Instituto de F\'{\i}sica,
Universidade Federal do Rio de Janeiro\\
C.P. 68528, 21941-972 Rio de Janeiro, Brazil}
\date{\today}
\begin{abstract}
We consider a new model of quantum walk on a one dimensional momentum space
that includes both discrete jumps and continuous drift.
Its time evolution has two stages; a Markov diffusion followed by localized dynamics. 
As in the well known quantum kicked rotor, this model can be mapped into a localized one-dimensional Anderson model.  
For exceptional (rational) values of its scale parameter,  the system exhibits resonant behavior and reduces to the usual discrete time quantum walk on the line. 
\end{abstract}
\end{frontmatter}

\section{Introduction}

The quantum random walk on the line has been studied as a natural
generalization of the classical random walk in relation with quantum
information processing. Two cases have been considered: the quantum walk with continuous
time or with discrete time steps \cite{Kempe}. One of the most striking
properties of the quantum random walk on the line is its ability to spread
over the line linearly in time, while its classical analog spreads out as the
square root of time. Experimental schemes to implement the quantum
walk have been proposed by a number of authors \cite{Dur,Travaglione,Sanders,Knight03a,Knight03b}.

Another subject that has drawn much attention in quantum information processing is dynamical localization (DL) \cite{Santos,Benenti,Terraneo,konno}, due to its potential for process control in a quantum computer.
DL is a quantum interference effect which suppresses quantum diffusion. It is typical in one dimensional periodically driven systems with chaotic classical counterparts, such as the quantum kicked rotor model
\cite{CCI79,Izrailev}. In recent works \cite{Georgeot,Levi}, quantum algorithms which
simulate the quantum kicked rotor faster than classical algorithms were
presented and we have proposed the use of these algorithms to describe the
evolution of discrete time quantum random walk \cite{Hadamard}.

In this work we consider, in section 2, a generalization of the quantum
random walk on a line,  along a line similar to the stroboscopic discrete--time quantum
walk presented in  \cite{Oliver,luczak}, is considered. In section 3, we numerically characterize its dynamical properties focusing in its localized and resonant behaviors. In section 4, we establish a connection with the one-dimensional  Anderson model from solid state physics and discuss similarities of the modified quantum walk with the kicked rotor model. The last section contains our conclusions.

\section{Modified quantum walk}

The discrete--time quantum walk on the line describes a particle which is free
to move over a lattice of interconnected sites labeled by an index $k$. In the
classical random walk, a coin flip randomly selects the direction of the
motion. In the quantum walk, the direction of the motion is selected by
introducing an additional degree of freedom, which we call the chirality,
which can take two values: ``left'' or ``right'', $|L\rangle$ or $|R\rangle$,
respectively. At each time step, a rotation (or, more generally, a unitary
transformation) of the chirality takes place and the particle moves according
to its final chirality state. The global Hilbert space of the system is the
tensor product $H_{s}\otimes H_{c}$ where $H_{s}$ is associated to the motion
on the line and $H_{c}$ is the chirality Hilbert space.

The conditional translation can be written as $S^{\sigma_{z}}$ where $S$ is
the one-site translation operator, $S|k\rangle=|k+1\rangle$, and $\sigma_{z}$
the usual $2\times2$ Pauli matrix. Then, this operator shifts the position on
the lattice left or right, according to the chirality component
\begin{align}
S^{\sigma_{z}}|L,k\rangle &  =|L,k+1\rangle\nonumber\\
S^{\sigma_{z}}|R,k\rangle &  =|R,k-1\rangle. \label{Ugen}%
\end{align}
In this work, the unitary operation on chirality is a Hadamard operation,
$H=(\sigma_{x}+\sigma_{z})/\sqrt{2}$. Then, the Hadamard walk on the line
evolves in a time step $\tau$ with $U=S^{\sigma_{z}}H$ as
\begin{equation}
|\Psi(t+\tau)\rangle=U|\Psi(t)\rangle. \label{evol1}%
\end{equation}

The sites $|k\rangle$ are usually associated with position eigenstates, but
they can equally well be interpreted as momentum eigenstates, i.e.
$P|k\rangle=\hbar k|k\rangle$, leading to a quantum walk in momentum space. In
fact, this is actually the case in proposed implementations of the quantum
walk using classical waves \cite{Knight03a,Knight03b}. Interpreting the lattice sites
as momentum eigenstates allows us to establish a connection with the
kicked rotator.

We modify the quantum walk described in eq.~(\ref{evol1}) by considering that
between two consecutive applications of $U$ the random walker drifts with
constant momentum. This drift is represented by a site-dependent phase shift
related to the kinetic energy $P^{2}/2m$, so that the new evolution takes the
form%
\begin{equation}
|\Psi(t+\tau)\rangle=e^{-i2\pi\Omega P^{2}/\hbar^{2}}U|\Psi(t)\rangle,
\label{newevol}%
\end{equation}
where $\Omega=\frac{\hbar\tau}{4\pi m}$ and $m$ is the particle mass. The wave
vector can be expressed as a spinor
\begin{equation}
\ket{\Psi(t)}=\sum\limits_{k=-\infty}^{\infty}
\left(
\begin{array}
[c]{c}
a_{k}(t)\\
b_{k}(t)
\end{array}
\right)  |k\rangle, \label{spinor}%
\end{equation}
with the upper (lower) associated to the left (right) chirality.
The unitary evolution implied by eq.(\ref{newevol}) can be written as the map
\begin{align}
a_{k}(t+\tau)  &  =\frac{1}{\sqrt{2}}\left[  a_{k+1}(t)\,+b_{k+1}(t)\,\right]
e^{-i2\pi\Omega k^{2}}\nonumber\\
b_{k}(t+\tau)  &  =\frac{1}{\sqrt{2}}\left[  a_{k-1}(t)\,-b_{k-1}(t)\right]
e^{-i2\pi\Omega k^{2}}. \label{mapa}%
\end{align}
Note that for integer values of $\Omega$ the usual Hadamard walk is
recovered.

The evolution of the momentum distribution, $F_k (t)\equiv\left|
a_{k}(t)\right|^{2}+\left| b_{k}(t)\right|^{2}$, is obtained from
eqs.(\ref{mapa}) as in our previous work \cite{Hadamard,master},
\begin{equation}
F_{k}(t+\tau)=\frac{1}{2}\left[  F_{k+1}(t)+F_{k-1}(t)\right]  +\beta
_{k+1}(t)-\beta_{k-1}(t), \label{probabilidad}%
\end{equation}
where $\beta_{k}(t)\equiv\Re\left[  a_{k}(t)b_{k}^{\ast}(t)\right]  $, is an
interference term with $\Re(z)$ indicating the real part of $z$. 
If these terms are neglected in eq.(\ref{probabilidad}), the resulting
evolution is markovian. In the continuum limit, taking $\tau =1$, the familiar diffusion
equation results
\begin{equation}
\frac{\partial F}{\partial t}=\frac{1}{2}\frac{\partial^{2}F}{\partial p^{2}}.
\label{difu}%
\end{equation}

The dynamics defined by eqs.~(\ref{mapa}) presents a very rich behavior, from quantum diffusion to quantum localization. For special values of $\Omega$ a resonant dynamics results, in which the variance of the distribution  $\sigma^{2}=\sum_{k}%
k^{2}F_{k}(t)-\left[  \sum_{k}k\,F_{k}(t)\right]  ^{2}$,  increases quadratically in time. In fact, as mentioned before, for integer $\Omega$ the usual quantum walk, eq.~(\ref{evol1}) is obtained. As is typical of near-resonant systems, small variations in the parameter $\Omega$ may lead to very different dynamics, as shown in Fig.~\ref{fig:loca}. 

\section{Localization and Resonances}
\begin{figure}[ptb]
\begin{center}
\includegraphics[scale=0.38]{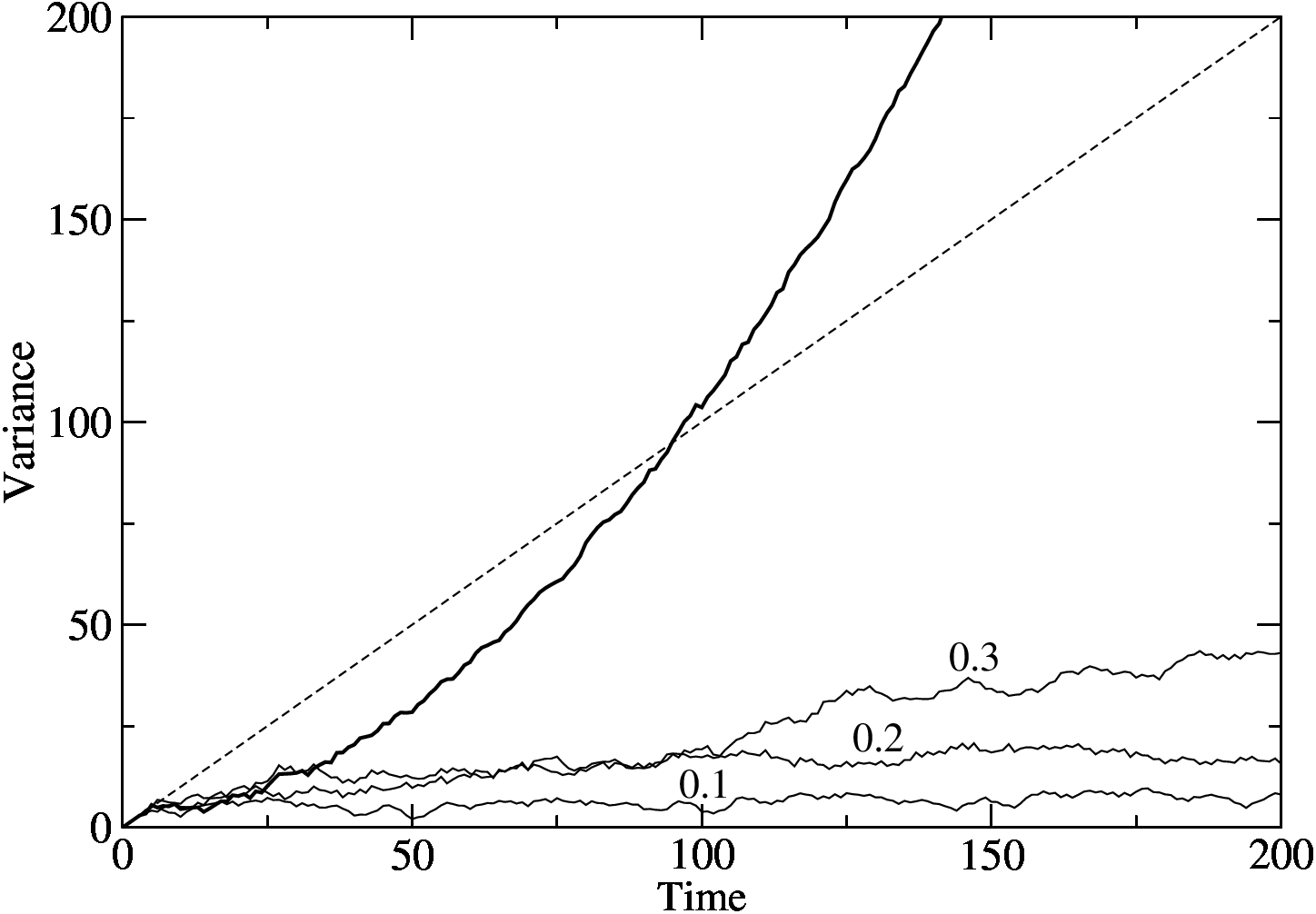}
\end{center}
\caption{Variance $\sigma^{2}$ as a function of time for several irrational values of the parameter $\Omega$, namely $2\pi\Omega=0.1,0.2,0.3$ (thin lines). The walker has been initialized at the origin in the state,
$|\Psi(0)\rangle=\frac{1}{\sqrt{2}}(1,i)|0\rangle$. The dashed line represents
the classical linear diffusion. The approximately parabolic full line corresponds to the rational case $\Omega=1/9.$}%
\label{fig:loca}%
\end{figure}
\begin{figure}[ht]
\centering
\includegraphics[scale=0.47]{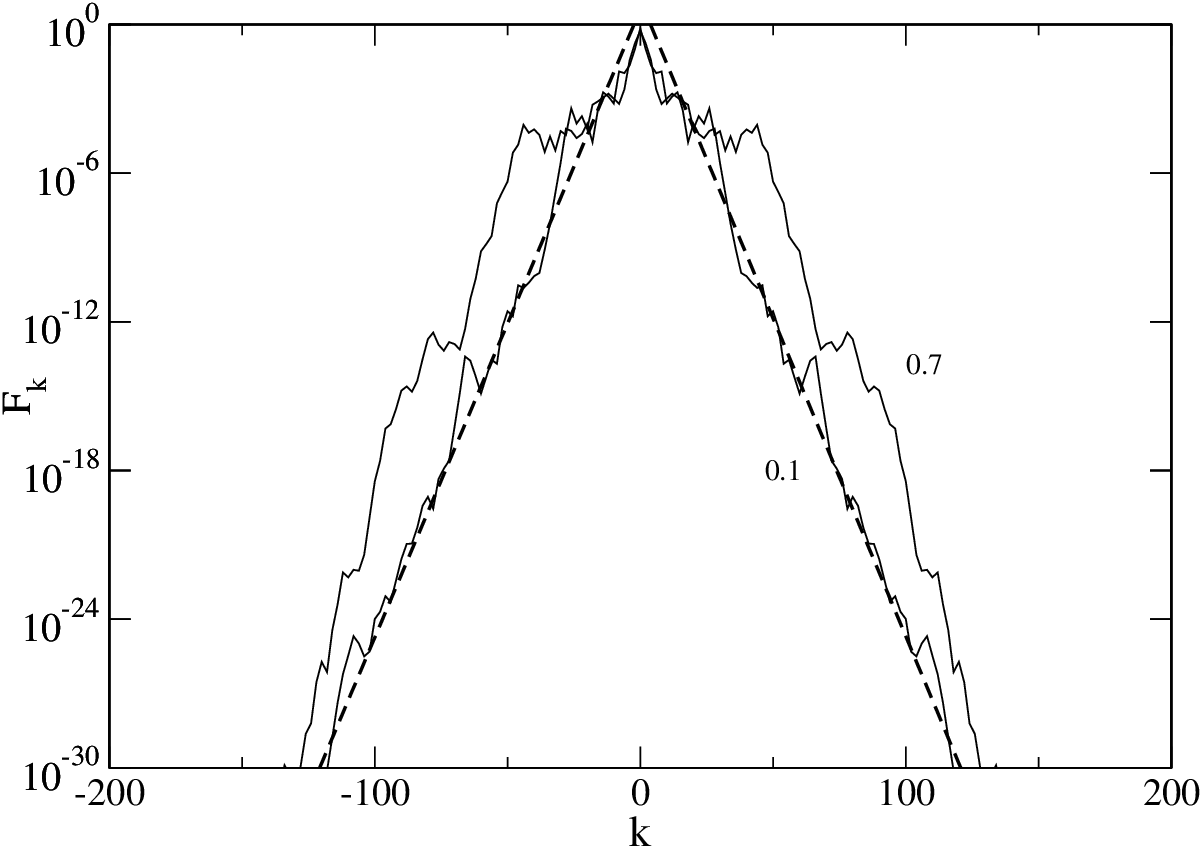}\caption{Momentum probability
distribution $F_{k}(t)$ at $t=2000$, as a function of the momentum index $k$
for  $2\pi\Omega=0.1,0.7$. The dashed line is the exponential fit for
$2\pi\Omega=0.1$  from where it is possible to obtain the localization length.
The initial condition is the same as in Fig.~\ref{fig:loca}. Notice that the 
vertical scale is logarithmic.}%
\vspace{0.80cm}
\label{fig:prob}%
\end{figure}

\vspace{0.8cm}
Two clear cut different behaviors are found depending on the rational or
irrational value of $\Omega.$ When $\Omega$ is irrational, an initial stage of
quantum diffusion with the diffusion coefficient expected from eq.~(\ref{difu}%
) is followed, after a characteristic time, by a localized dynamics in which
the growth ceases due to quantum interference (see Fig.~\ref{fig:loca}). The
exponential character of this localization is apparent from
Fig.~\ref{fig:prob}, where the distribution $F_{k}(t)$ is shown after $2000$
time steps, when DL has already set in.

\begin{figure}[ht]
\centering
\includegraphics[scale=0.63]{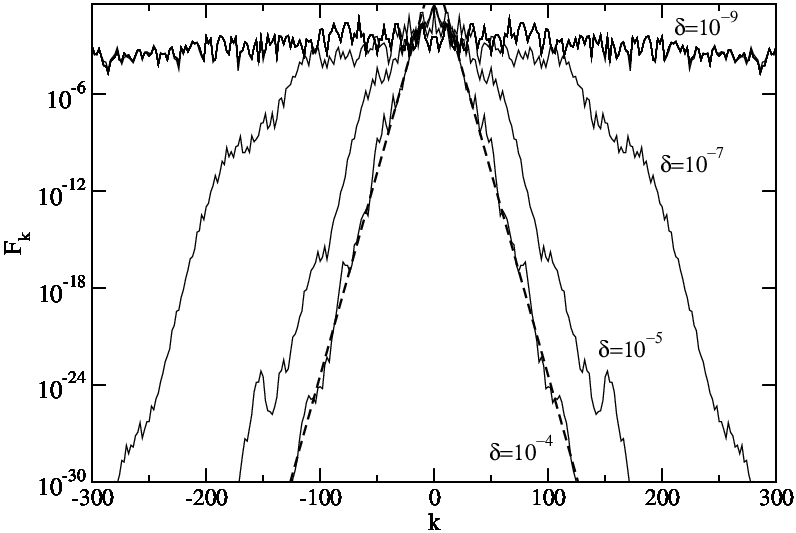}\caption{Momentum probability
distribution $F_{k}(t)$ at $t=2000$, in logarithmic scale, as a function of
the momentum index $k$ for $\Omega=\frac{1}{11}+\delta\ $, with $\delta=10^{-4}%
,10^{-5},10^{-6}$and $10^{-9}.$ The dashed line is the exponential fit for
$\delta=10^{-4}$. The initial condition is the same as in Fig.~\ref{fig:loca}.}
\label{fig:prob2}
\end{figure}

In Fig.~\ref{fig:prob2} the loss of localization that takes place for
rational values of $\Omega$ is illustrated by considering a set of near-resonant parameter $\Omega=\Omega_r+\delta$ with 
$\Omega_r$ rational and $\delta~\ll~1$. For not too small values of $\delta$ ($\Omega$ approximately irrational) the distribution is exponentially localized. As $\delta$ is gradually decreased the exponential localization transforms into an almost uniform distribution.  To aid visualization we have plotted only the non-zero probabilities in
the distributions since the map~(\ref{mapa}) implies that if a single site is initially occupied, at any later time only odd or even sites are occupied with non-zero probabilities. 

\begin{figure}[ptbh]
\begin{center}
\includegraphics[scale=0.4]{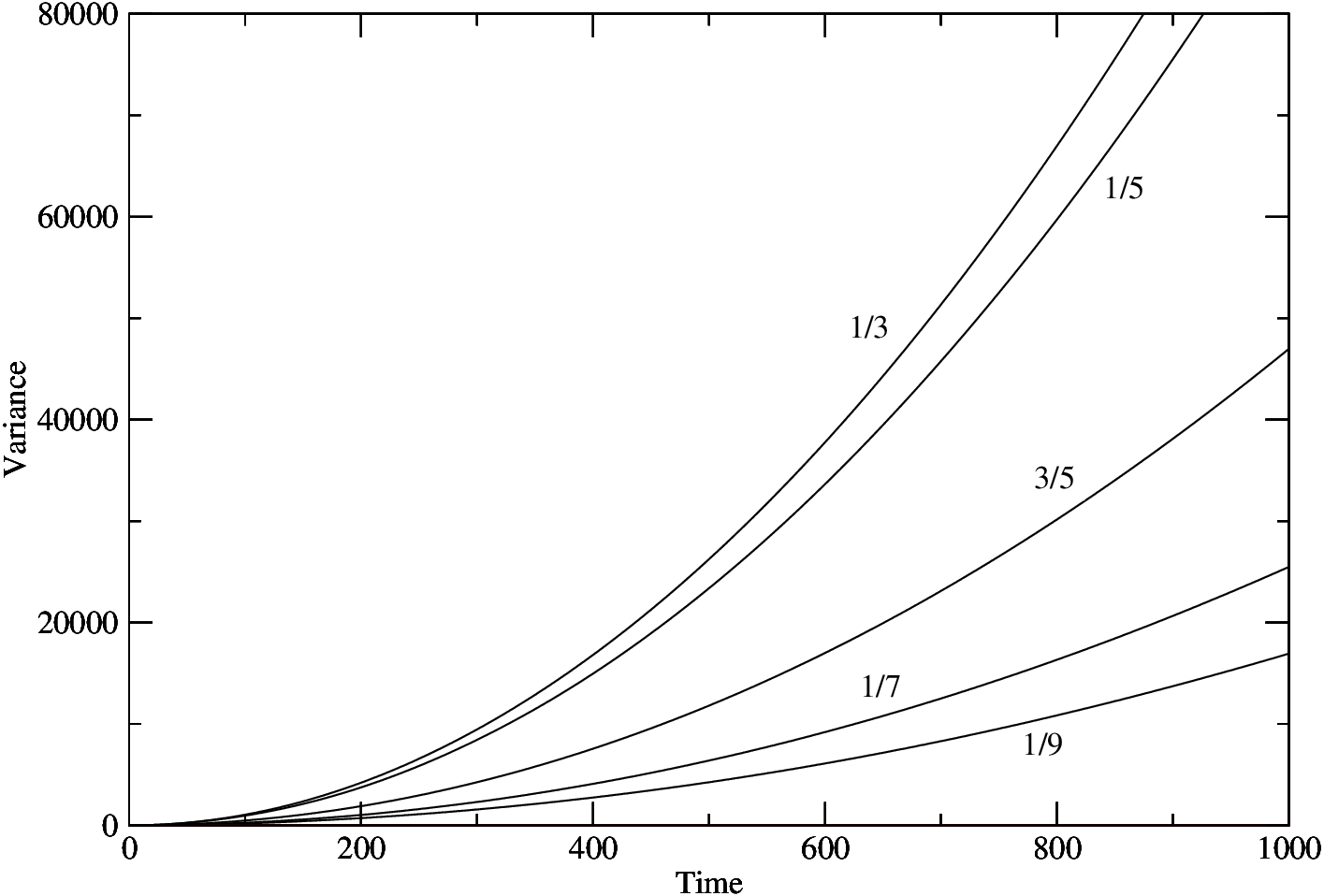}
\end{center}
\caption{Quadratic growth of the variance as a function of time for several
rational values of the parameter $\Omega=1/3,1/5,3/5,1/7,1/9$.}
\label{fig:reso}
\end{figure}

\vspace{0.8cm}
When $\Omega$ is rational, the variance increases quadratically in time, as shown in
Figs.~\ref{fig:loca} and \ref{fig:reso}. In this case, the characteristic transition frequency
between momentum eigenstates is commensurable with the random walk's 
frequency. This is the condition for resonant behavior. To understand such a
resonant regime, it is possible to use the theoretical developments presented
either in our recent work \cite{Hadamard} where we consider the principal
resonances ($\Omega$ integer) of the quantum random walk or, alternatively, in
the detailed results obtained for a related system, the kicked rotor, in the past \cite{Izrailev}.
\vspace{0.8cm}
\vspace{0.8cm}
If $\Omega$ is a non-integer rational number $\Omega=\frac{p}{q}$, we are in the presence of secondary resonances. 
The variance still displays a quadratic growth but it is slower than the growth
associated to the main resonances. This is due to the fact that while in the
main resonances all phases add constructively, in secondary resonances 
there are (at most $q$) different phases that repeat themselves periodically in
time. These phases determine the time required for coherent behavior to influence the dynamics.\\
\\
\begin{figure}[h]
\begin{center}
\includegraphics[scale=0.6]{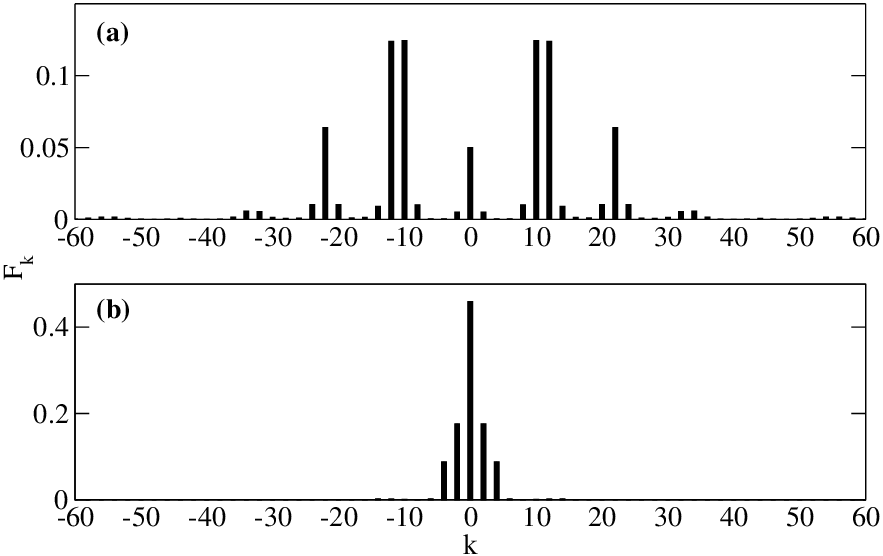}
\end{center}
\caption{In the upper frame, momentum probability distribution $F_k$ as a
function of the momentum $k$ for $\Omega=\frac{1}{11}$. In the lower frame the
same distribution for $\Omega=\frac{1}{11}+10^{-4}$. In both frames the
distribution is calculated for t=$2000\tau$.}
\label{fig:ldos}
\end{figure}
\\
In Fig.~\ref{fig:ldos} (a),  the probability distribution after $2000$ steps of the secondary resonance $\Omega=\frac{1}{11}$ is shown. A constructive interference phenomenon can be clearly seen when the 
momenta are multiples of the denominator $q=11$, \textit{i.e.~}
$k~=~...,-22,-11,0,11,22,...$. On the other hand, Fig.~\ref{fig:ldos} (b)  shows that the
distribution for $\Omega=\frac{1}{11}+10^{-4}$, after the same time number of steps,
is clearly localized. Thus there is a high sensitivity of this system to small
variations of $\Omega$ near a rational value. Although $\Omega$ is strictly rational in both cases,  for the second, the time at which the system becomes sensitive to its rational character is much greater than the time considered in the figure.

\section{Connection with Anderson localization}

As we have shown, the map~(\ref{mapa}) generates the
same qualitative behavior as that of the quantum kicked rotor. This last
system has been extensively studied because its evolution operator can be expressed analytically and its classical version
(the standard or Chirikov map) is a cornerstone in the theory of chaotic
Hamiltonian systems \cite{Ott}. In particular, the phenomena of DL was first
observed in numerical experiments \cite{CCI79} for the quantum kicked rotor. In
the last decade, direct observations of DL have been done using samples of cold atoms
interacting with far-detuned light \cite{Moore}. When the light field is switched on and off periodically, 
the resulting Hamiltonian is essentially that of the quantum kicked rotor \cite{expqkr1,expqkr2,Ammann}. 
In a particularly interesting theoretical development \cite{Grempel1,Grempel2}, the kicked rotor Hamiltonian has been mapped into an Anderson model in a one-dimensional lattice \cite{Anderson1,Anderson2}. This connected the dynamical exponential localization in momentum space to the spatially localized wave functions of interest in solid state physics. 

There is a strong similarity between the modified quantum walk and the kicked rotor. The Floquet (\textit{i.e.} one-step evolution) operator of the kicked rotor involves a kick operator $e^{-iK\cos\theta}$, with $K$ determining the kick strength and $\theta$ the angular position of the rotor,  followed by a free evolution over a time step, $e^{-iP^{2}/2\hbar}$.  For the modified
quantum walk, the kick operator is replaced by the conditional translation and the Hadamard rotation $U=S^{\sigma_{z}}H$. 

Not surprisingly, the modified quantum walk on the line can also be mapped into a localized one dimensional
Anderson equation. Thus, this momentum localization may also be thought of as a dynamical form of Anderson localization. 
In order to show this, we relate the Floquet eigenstates of the modified walk with the exponentially localized states of the Anderson model. 

The characteristics of the modified quantum walk imply that the Hamiltonian of
the system is periodic in time and Floquet theory \cite{Haake,Reichl} applies. 
Consider a Floquet eigenstate $\ket{\Psi_w}$ of the one-step operator $e^{-i2\pi\Omega P^{2}/\hbar^{2}}\,U$, so that $e^{-i2\pi\Omega P^{2}/\hbar^{2}}\,U\ket{\Psi_w}=e^{-iw}\ket{\Psi_w}$ and  $w$ is the associated quasienergy. Then $\ket{\Psi_w}=e^{-iwt}\ket{\Phi_{w}(t)}$ with $\ket{\Phi_{w}(t+\tau)}=\ket{\Phi_{w}(t)}$. The time dependence of Floquet states is trivial and we shall suppress it from our notation.  
In the case of Floquet eigenstates, the map (\ref{mapa}) reduces to the set of equations
\begin{align}
f_{k}a_{k} &  =  a_{k+1}\,+b_{k+1}  \nonumber\\
f_{k}b_{k} &  =  a_{k-1}\,-b_{k-1} \label{mapa2},
\end{align}
where $f_{k}=\sqrt{2}e^{i(2\pi\Omega k^{2}-w)}$. Note the dependence on the quasienergy $w$ implicit in (\ref{mapa2}). After decoupling these equations and performing the change of variable
\begin{equation}
\binom{\alpha_{k}}{\beta_{k}}=i^{k}f_{k}\binom{a_{k}}{\,b_{k}},
\end{equation}
we obtain
\begin{align}
g_{k}\alpha_{k}&= \alpha_{k+1}\,+\alpha_{k-1} \nonumber\\
\tilde g_{k}\beta_{k}&= \beta_{k+1}\,+\beta_{k-1},
\label{mapa4}
\end{align}
where we have defined $g_{k}\equiv i(f_{k+1}-f_{k}^{\ast})$ and $\tilde g_{k}\equiv i(f_{k-1}-f_{k}^{\ast})$. 

These equations strongly suggest a relation with Lloyd's model \cite{lloyd}, a tight binding model of an electron on a one-dimensional
disordered lattice. This model is a special case of the one-dimensional Anderson model \cite{Anderson2} in the case of nearest neighbor interaction, which has been analytically solved \cite{Reichl}. The difference with our model is that $g_{k}$ and $\tilde g_{k}$ in eqs.~(\ref{mapa4}) have a non-zero imaginary part. 

It is clear that the equations for both spinor components have same nature, so we focus on the first and show that it can be transformed into an Anderson equation
\begin{equation}
T_{k}\alpha_{k}+\sum\limits_{l\neq k}W_{kl}\alpha_{l}=0,\label{Anderson}
\end{equation}
where $T_{k}$ is the kinetic energy at site $k$ and $W_{kl}$ is the hopping
term. It is well established \cite{Reichl} that if the $T_{k}$ are pseudo-random functions of
the site $k$ and the hopping term $W_{km}$ decays sufficiently fast with
distance $|k-m|,$ the wavefunction is exponentially localized. We use
eq.~(\ref{mapa4}) recursively to express it as an equation with real coefficients of the form
(\ref{Anderson}) with 
\begin{equation}
T_{k}=g^i_k\,[g_{k-1}^i+g^i_{k+1}]+g^i_{k+1}g^i_{k-1}\left[(g^r_{k})^2+(g^i_k)^2\right],\label{cinetica}
\end{equation}
$g^r_k$ and $g^i_k$ are the real and imaginary parts of $g_k$.  

The potential term $W_{kl}$ is
\begin{equation}
W_{kl}=\left\{
\begin{array}{cr}
g^{i}_{l-2}\,g^i_{l-3}& l=k+2\\
-g^i_{l-2}\left[g^r_{l}g^i_{l-1}+g^i_{l}g^r_{l-1}\right]& l=k+1\\
-g^i_{l+2}\left[g^r_{l}g^i_{l+1}+g^i_{l}g^r_{l+1}\right]& l=k-1\\
g^i_{l+2}\,g^i_{l+3}&l=k-2\\
0&\mbox{other cases}.
\end{array}
\right.  \label{potencial}%
\end{equation}

For irrational $\Omega$, the kinetic term $T_{k}$ has
a pseudo-random behavior in $k$ and we find that its probability distribution is narrowly peaked as required for localization. The hopping term now couples each
component of the eigenstate with its two closest neighbors. 
An equation similar to eq.~(\ref{Anderson}) holds for the other chirality component.  
We have shown, therefore, that the modified random walk can be mapped into a localized one-dimensional Anderson model. 

\section{Conclusions}

A modified version of the discrete time quantum walk on the line has been considered. 
After an initial diffusive stage, dynamical localization takes place for most values of the parameter
$\Omega$. However, for rational values of the parameter the dynamics of the
system corresponds to a resonant behavior. In particular the usual quantum
walk is recovered when $\Omega$ is an integer. The modified walk does not show
a quadratic increase of its variance for all values of the parameter; rather,
this happens only for exceptional (rational) values, for which the quantum
spreading is faster than the classical diffusion spread.

The modified quantum walk has been mapped into a one-dimensional Anderson model, as was previously done in the
case of the kicked rotor \cite{Grempel1,Grempel2}. The similarity between the modified quantum walk and the kicked rotor has thus been established both numerically and analytically. Both models are characterized by two parameters, a strength parameter and a scale parameter. In the case of the modified quantum walk, the scale parameter is $\Omega=\frac{\hbar\tau}{4\pi m}$ . The strength parameter can be associated to the parameter determining the generalized rotation of chirality, as discussed in \cite{Hadamard}.

In the generic case, when $\Omega$ is irrational, a new kind of quantum random
walk on the line, with exponentially localized wave functions, is obtained.
This feature opens interesting possibilities for quantum information
processing. In particular the question arises, may DL be used for controlling the spread of a wavefunction? Furthermore, since DL states are robust against moderate
decoherence \cite{Terraneo,Ammann}, can they be used in preserving quantum
information against environment induced noise? It would be interesting to
experimentally observe the effects of resonances and DL when one of the
existing experimental proposals for the quantum walk \cite{Dur,Travaglione} is
realized. We note that these effects might also be observed in the context of
classical wave simulations for the quantum walk \cite{Knight03a,Knight03b}.

We acknowledge support from PEDECIBA and PDT S/C/OP/29/84. R.D.
acknowledges partial financial support from the Brazilian National Research
Council (CNPq) and FAPERJ (Brazil). A.R. acknowledges useful comments made by V. Micenmacher.

\end{document}